\documentclass[aps,prl,twocolumn,twoside,superscriptaddress,nofootinbib,preprintnumbers]{revtex4-1}

\usepackage[bookmarks=false]{hyperref}
\usepackage{graphicx}
\usepackage{amsmath}
\usepackage{mathrsfs}
\usepackage{xspace}
\usepackage{color}
\usepackage[normalem]{ulem}
\usepackage[abs]{overpic}
\usepackage{comment}

\newcommand{\eq}[1]{Eq.~\eqref{eq:#1}}

\newcommand{\fig}[1]{Fig.~\ref{fig:#1}}

\newcommand{\df}{\mathrm{d}}
\newcommand{\img}{\mathrm{i}}
\newcommand{\sdt}{\!\cdot\!}

\newcommand{\al}{\alpha}

\newcommand{\ga}{\gamma}

\newcommand{\de}{\delta}

\newcommand{\si}{\sigma}

\newcommand{\w}{\omega}

\newcommand{\cH}{\mathcal{H}}
\newcommand{\cJ}{\mathscr{J}}

\newcommand{\bn}{\bar{n}}

\newcommand{\one}{{(1)}}
\newcommand{\two}{{(2)}}

\newcommand{\nn}{\nonumber}

\newcommand{\MCFM}{\textsc{MCFM}\xspace}
\newcommand{\Pythia}{\textsc{Pythia}\xspace}

\allowdisplaybreaks[2]

\begin{document}


\preprint{\vbox{\hbox{CERN-TH-2020-078}}}
\preprint{\vbox{\hbox{Nikhef 20-008}}}

\title{Recoil-free azimuthal angle for precision boson-jet correlation}

\author{Yang-Ting Chien}
\email{yang-ting.chien@stonybrook.edu}
\affiliation{C.N.~Yang Institute for Theoretical Physics, Stony Brook University, Stony Brook, NY 11794, USA}
\affiliation{Center for Frontiers in Nuclear Science, Stony Brook University, Stony Brook, NY 11794, USA}

\author{Rudi Rahn}
\email{rudi.rahn@uva.nl}
\affiliation{Nikhef, Theory Group,
	Science Park 105, 1098 XG, Amsterdam, The Netherlands}
\affiliation{Institute for Theoretical Physics Amsterdam and Delta Institute for Theoretical Physics, University of Amsterdam, Science Park 904, 1098 XH Amsterdam, The Netherlands}

\author{Solange Schrijnder van~Velzen}
\email{s.v.schrijndervanvelzen@uva.nl}
\affiliation{Nikhef, Theory Group,
	Science Park 105, 1098 XG, Amsterdam, The Netherlands}
\affiliation{Institute for Theoretical Physics Amsterdam and Delta Institute for Theoretical Physics, University of Amsterdam, Science Park 904, 1098 XH Amsterdam, The Netherlands}

\author{Ding Yu Shao}
\email{dingyu.shao@cern.ch}
\affiliation{Center for Frontiers in Nuclear Science, Stony Brook University, Stony Brook, NY 11794, USA}
\affiliation{Department of Physics and Astronomy, University of California, Los Angeles, CA 90095, USA}
\affiliation{Mani L. Bhaumik Institute for Theoretical Physics, University of California, Los Angeles, CA 90095, USA}
\affiliation{Department of Physics and Center for Field Theory and Particle Physics, Fudan University, Shanghai, 200433, China}
\affiliation{Key Laboratory of Nuclear Physics and Ion-beam Application (MOE), Fudan University, Shanghai, 200433, China}

\author{Wouter J.~Waalewijn}
\email{w.j.waalewijn@uva.nl}
\affiliation{Nikhef, Theory Group,
	Science Park 105, 1098 XG, Amsterdam, The Netherlands}
\affiliation{Institute for Theoretical Physics Amsterdam and Delta Institute for Theoretical Physics, University of Amsterdam, Science Park 904, 1098 XH Amsterdam, The Netherlands}

\author{Bin Wu}
\email{b.wu@cern.ch}
\affiliation{CERN, Theoretical Physics Department, CH-1211, Geneva 23, Switzerland}

\begin{abstract}
The azimuthal decorrelation between a vector boson and a jet is an essential hard probe in high energy proton-proton and heavy-ion collisions. We overcome intrinsic limitations of previous studies by using a recoil-free axis, achieving unprecedented next-to-next-to-leading logarithmic accuracy with small  nonperturbative corrections. This choice of axis also makes the observable robust in the presence of a large background. Furthermore, the azimuthal angle distribution is minimally changed when determined using only charged particle tracks, which offer superior angular resolution for precise measurements. Our effective field theory includes full jet dynamics, and we find contributions from linearly-polarized gluon transverse momentum distributions in the initial \emph{and} final state.
\end{abstract}

\maketitle

{\it Introduction. --}
The 
production of a vector boson ($V$) in association with jets is a crucial process in $pp$ and heavy ion
collisions. It is an important background in the study of Standard Model processes  (e.g.~to control b-tagging for $t \bar t$
measurements~\cite{Abe:1995hr}) and the search for physics beyond the Standard Model (see \cite{Blumenschein:2018gtm} for a recent review), and a prime channel to study the effects of
the quark-gluon plasma produced in heavy-ion collisions~\cite{Kartvelishvili:1995fr, Sirunyan:2017jic, Cao:2020wlm}. The precise theoretical prediction for such processes relies on advances in both fixed-order calculations and all-order resummation of large (Sudakov) logarithms. In $pp$ collisions, the fixed-order calculations of such processes have  reached next-to-next-to-leading order in QCD \cite{Ridder:2015dxa,Boughezal:2015dva,Boughezal:2015ded,Campbell:2016lzl,Gehrmann-DeRidder:2017mvr,Chen:2019zmr}, while in the back-to-back limit the Sudakov logarithms in the total transverse-momentum distribution of $V$+jet  have only been resummed up to next-to-leading logarithmic (NLL) accuracy~\cite{Chen:2018fqu,Sun:2018icb,Chien:2019gyf}. The relatively large uncertainties in the resummed result at NLL accuracy (see the discussion in \cite{Chien:2019gyf}) is one of the main obstacles to a precise prediction for such processes. In order to match to high-accuracy fixed-order calculations, one has to extend these resummation techniques to higher order. This, however, has been hamstrung by their intrinsic limitiations, such as the presence of non-global logarithms~\cite{Dasgupta:2001sh}.

In this letter, by explicitly calculating the angular decorrelation of $Z+$jet~\cite{Chatrchyan:2013tna,Khachatryan:2016crw,Sirunyan:2017jic,Aaboud:2017kff} in the back-to-back limit up to next-to-next-to-leading logarithms (NNLL), we show that one can overcome these limitations by using two main ingredients: First, a recoil-free jet axis, obtained using the Winner-Takes-All (WTA) recombination scheme~\cite{Salam:WTAUnpublished,Bertolini:2013iqa}.
And second, factorization-based methods from Soft-Collinear Effective Theory
(SCET)~\cite{Bauer:2000yr,Bauer:2001ct,Bauer:2001yt,Bauer:2002nz,Beneke:2002ph}, which
allow us to obtain extremely precise predictions for the angular decorrelation.
We also find --- for the first time --- that predictions for this observable include linearly-polarized gluon
transverse-momentum-distributions (TMDs) in the initial \emph{and} final state. 
For the initial state this novel mode of appearance arises from spin superpositions for
one gluon \cite{Boer:2009nc} (for Higgs production this arises instead from spin interference
between multiple initial-state gluons~\cite{Catani:2010pd}).

\begin{figure}
\includegraphics[width=0.4\textwidth]{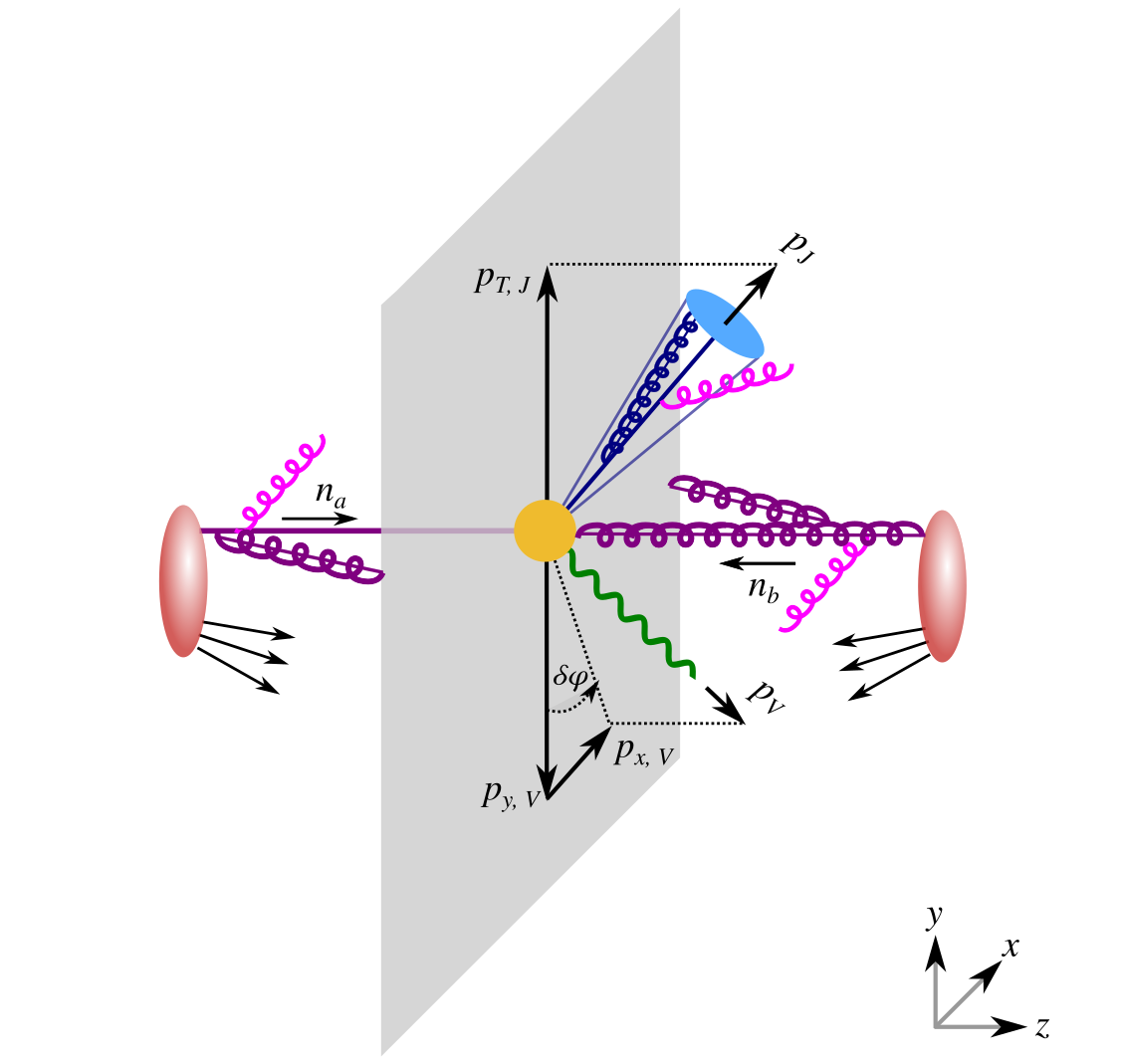}
\caption{The azimuthal angle between the vector boson (green) and WTA jet axis (blue) is related to the  momentum of the vector boson $p_{x,V}$ transverse to the colliding protons (red) and jet. Collinear initial  (purple) and final-state (blue) radiation and soft radiation (pink) is also shown.
\label{fig:coordinate}}
\end{figure}

The recoil-free axis is the prime driver of success for this letter: Its unique
ability to separate the effects of soft and collinear radiation reduces the
impact of soft recoil~\cite{Larkoski:2014uqa},
removing the otherwise problematic non-global logarithms.
This allows us to make predictions 
for very small angular
deviations from --- at leading order --- the back-to-back case (see \fig{coordinate}). 
Reduced soft sensitivity is also crucial in environments with a lot of contamination, 
such as high-energy nuclear collisions. 
The use of a recoil-free axis in lepton-ion collisions
was proposed in~\cite{Arratia:2019vju}, where the simpler color and spin
structure removes the linearly-polarized contributions.
Note that using the WTA axis does not require a complete new calibration of the jet. Rather, one can recluster jets to obtain the WTA axis, which for the proposed measurement has a negligible impact.

Furthermore, we show from first principles that measurements using tracks have
an almost identical distribution, implemented using track functions~\cite{Chang:2013rca,Chang:2013iba}.
This makes it possible to exploit the superior angular resolution of the tracking system (compared to calorimetry).

{\it Factorization. --}
We start by giving a precise definition of our observable.
The azimuthal angle $\Delta \phi$ between the vector boson ($V$) and the jet
($J$) is directly related to the momentum component $p_{x,V}$ of the vector
boson, which
is perpendicular to the plane defined by the colliding beams (labelled $a$, $b$)
and the jet axis, see \fig{coordinate}. Explicitly, $\pi - \Delta \phi \equiv \de \phi \approx \sin(\de \phi) =
|p_{x,V}|/p_{T,V}$, where we introduce $\de \phi$ such that the back-to-back limit 
corresponds to $\de \phi \to 0$. (This is similar to the planar limit for $e^+e^- \to 3$ jets investigated in \cite{Arpino:2019ozn}, which is not a transverse momentum observable.)

We now present the factorization formula, assuming $\delta \phi \ll R$ (other cases were considered in \cite{Gutierrez-Reyes:2018qez}). Since the azimuthal angle is related to $p_{x,V}$, we consider momentum conservation along this direction. In the infrared limit $\de \phi \to 0$,
\begin{align} \label{eq:mom}
p_{x,a} + p_{x,b} + p_{x,c} + p_{x,S} + p_{x,V} = 0
\,,\end{align}
where $p_{x,a}, p_{x,b}$ ($p_{x,S}$) originate from collinear (soft)
initial- (initial- and final-) state radiation. A non-zero contribution
$p_{x,c}$ from collinear radiation inside the jet arises because the jet momentum and axis are \emph{not} aligned for the WTA axis~\cite{Bertolini:2013iqa}.
Eq.~\eqref{eq:mom} trivially gives access to our observable since $p_{x,J}=0$ due to our choice of coordinate system.
Writing this in terms of the Fourier conjugate variable $b_x$, we follow the standard steps~\cite{Stewart:notes,Becher:2014oda,Schwartz:2013pla} in SCET to
obtain the following factorization formula~\cite{longpaper}
\begin{align} \label{eq:fact}
   &\frac{\df \si}{\df p_{x,V}\, \df p_{T,J}\, \df y_V\, \df \eta_J} 
   \\
   &= \!\int\! \frac{\df b_x}{2\pi}\, e^{\img p_{x,V} b_x} \sum_{i,j,k} B_i(x_a,b_x)  B_j(x_b,b_x)  S_{ijk}(b_x, \eta_J)
   \nn \\ & \quad \times
   \cH_{ij \to Vk}(p_{T,V}, y_V - \eta_J) 
   \cJ_k(b_x)  \Bigl[1 + \mathcal{O}\Bigl(\frac{p_{x,V}^2}{p_{T,V}^2}\Bigr)\Bigr],
\nn\end{align}
where we suppressed the dependence on the renormalization scales. 
Here, $y_V$ ($\eta_J$) denote the (pseudo)rapidity of the vector boson (jet),
the sum on $i,j,k$ runs over the partonic channels (including linearly-polarized
gluon beam and jet functions), and the hard function $\cH_{ij \to Vk}$ describes
the short-distance scattering. The contribution to $p_{x,V}$ from collinear
initial- and final-state radiation and soft radiation is encoded in the
(standard) TMD beam functions
$B_{i,j}$, the TMD jet function $\cJ_k$ and the soft function $S_{ijk}$, which will be discussed in more detail below. 

We now comment on the crucial role played by the WTA recombination scheme in deriving the factorization in \eq{fact}. By using a recoil-free jet axis, 
the effect of soft radiation is power suppressed in the jet algorithm. Consequently,  soft radiation is not treated differently inside or outside the jet and only leads to the total recoil in \eq{mom}~\cite{Gutierrez-Reyes:2018qez}. This allows us to overcome the intrinsic limitation posed by NGLs~\cite{Dasgupta:2001sh} and extend the accuracy beyond the NLL results of \cite{Banfi:2008qs},  which used jets obtained with a $p_T^n$-weighted recombination scheme with $n=1$ (for $n>1$ the effect of soft radiation on the axis is power suppressed, and the WTA scheme corresponds to $n \to \infty$).
Since the cross section is independent of the renormalization scale, an important check of \eq{fact} is that the anomalous dimensions of the ingredients cancel against each other, which we have verified up to two loops.
In principle, \eq{fact} may receive corrections from factorization violating effects~\cite{Collins:2007nk,Rogers:2010dm,Catani:2011st,Forshaw:2012bi}, which can be systematically accounted for in SCET using a Glauber mode~\cite{Rothstein:2016bsq}, and are not required at the level of accuracy we consider.

{\it Resummation. --}
\eq{fact} enables the resummation of large logarithms by separating the physics at different scales. Specifically, we evaluate each ingredient at its natural scale, and use the ingredients' renormalization group equations to evolve them to a common scale, thereby resumming the logarithms of $\delta \phi$.
 The collinear (beam and jet) and soft ingredients in \eq{fact} have the same virtuality and are only separated in rapidity. This requires a rapidity regulator, for which we adopt the $\eta$-regulator~\cite{Chiu:2011qc,Chiu:2012ir}, leading to rapidity divergences of $1/\eta$ and a corresponding evolution in the rapidity renormalization scale $\nu$ that sums (large) rapidity logarithms. (For other choices of rapidity regulators, see e.g.~\cite{Ji:2004wu,Chiu:2009yx,Becher:2010tm,Collins:2011zzd,Becher:2011dz,GarciaEchevarria:2011rb,Li:2016axz}.) The natural scales of the ingredients in \eq{fact} are:
\begin{align}
  \mu_H &\sim \nu_B \sim \nu_J \sim p_{T,V} \sim m_V\,,  \nn \\
  \mu_B &\sim \mu_J \sim \mu_S \sim \nu_S \sim 1/|b_x|\,.
\end{align}
In this letter we will present numerical results at NNLL accuracy, which requires the ingredients in \eq{fact} at one-loop order, their anomalous dimensions at two-loop order~\cite{Moch:2005tm,Moch:2005id,Becher:2009cu,Gehrmann:2014yya,Echevarria:2015byo,Luebbert:2016itl} and the cusp anomalous dimension at three-loop order~\cite{Moch:2004pa,Moch:2005tm}. We note that the anomalous dimensions for the linearly-polarized beam and jet functions are the same as their unpolarized counterparts. Furthermore, most of the ingredients for N$^3$LL resummation are available. 

{\it Ingredients. --}
We now will describe the various ingredients entering in \eq{fact}.
For our NNLL predictions we need the hard function at one-loop order~\cite{Arnold:1988dp,Becher:2012xr}, and a new contribution multiplying \emph{linearly-polarized} gluon
beam~\cite{Tackmann} and jet functions that we calculate here:
\begin{align}
    \cH^L_{ij\to Vk} = \frac{x_a x_b p_T}{8\pi \hat{s}^2}\bigl|\overline{M}^L(ij\to Vk)\bigr|^2
\end{align}
with
\begin{align} \label{eq:H}
   \bigl|\overline{M}^L(qg\to \gamma q)\bigr|^2&=  \frac{32\pi^2 \alpha_{em} \alpha_s e_q^2}{N_c} \frac{\hat u m_V^2}{\hat s \hat t} , \\
   \bigl|\overline{M}^L(q\bar{q}\to \gamma g)\bigr|^2& = - \frac{32\pi^2 \alpha_{em} \alpha_s e_q^2(N_c^2-1)}{N_c^2} \frac{\hat s m_V^2}{\hat u \hat t}
\,.\nn\end{align}
Here $\hat s, \hat t, \hat u$ are the partonic Mandelstam variables, and $m_V^2$ is the off-shellness of the photon. For the $Z$ boson we have the usual replacement of the coupling.

In \eq{fact} $\cH^L$ gets accompanied by one linearly-polarized gluon beam or jet function. Since these start at order $\al_s$, we only need the LO results for $\cH^L$.
Interestingly, the linearly-polarized contributions enter the cross section already at NLO, instead of NNLO for Higgs production~\cite{Mantry:2009qz,Catani:2010pd,Luo:2019bmw,Gutierrez-Reyes:2019rug}. This is also the first time a linearly-polarized \emph{jet} function appears, which we discuss in more detail below.

\begin{figure*}[t]
\hfill \includegraphics[width=0.4\textwidth]{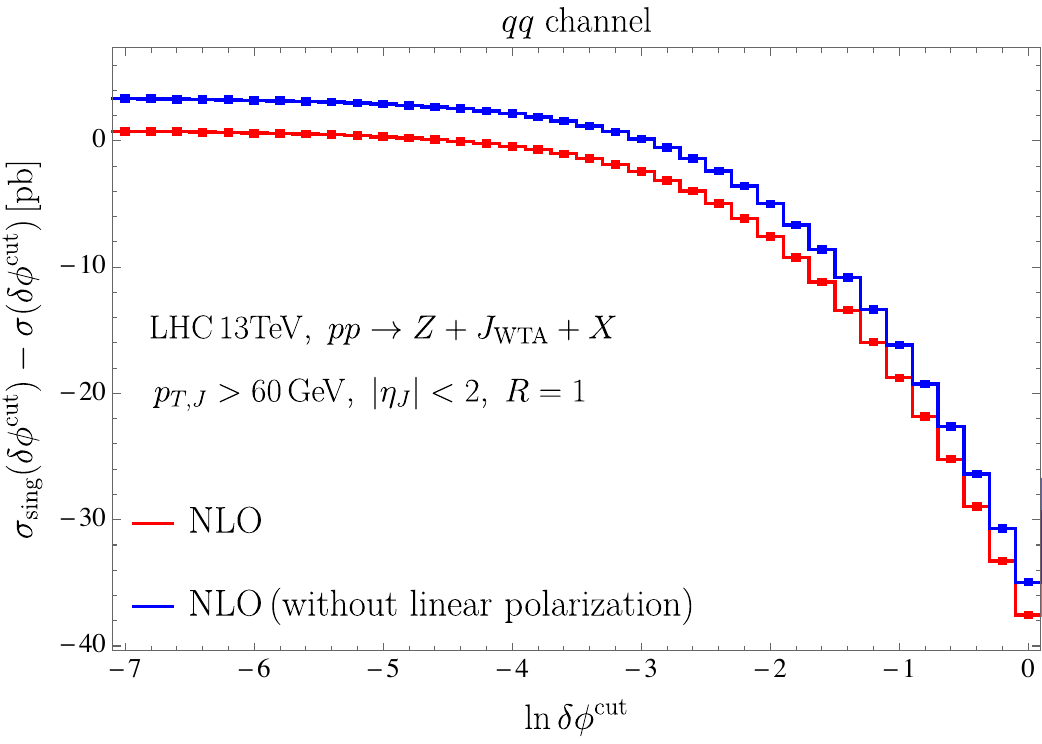}\hfill\hfill\hfill
\includegraphics[width=0.4\textwidth]{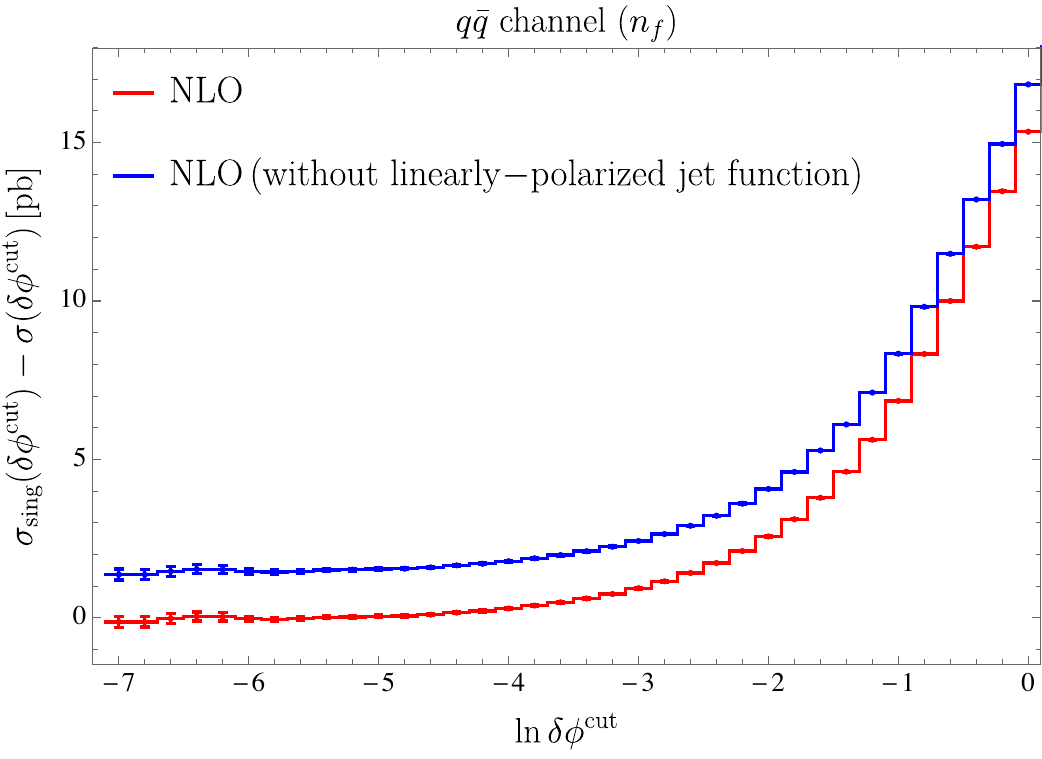} \hfill \phantom{p}
\caption{Difference between the singular cross section in \eq{fact} and the cross section from \MCFM at NLO with a cut $\de \phi<\de \phi^{\rm cut}$. Shown are the contribution from $qq$ (left) and $q\bar q$ (right) PDF flavors. Jets are identified using the anti-$k_T$ algorithm with $R=1$ and the leading jet fulfills $p_{T,J}>60\,{\rm GeV}$ and $|\eta_J|<2$. In the right plot we only consider NLO corrections proportional to $n_f$. 
\label{fig:mcfm}}
\end{figure*} 

Up to order $\al_s^2$, the soft function $S_{ijk}$ can be determined from the standard TMD soft function $S$~\cite{Echevarria:2015byo,Luebbert:2016itl,Li:2016ctv}. For exchanges involving only two Wilson lines, we can perform a boost to make them back-to-back. Similar to \cite{Gao:2019ojf}, our observable is perpendicular to the boost, so only the rapidity regulator is affected by such a boost (see e.g.~\cite{Kasemets:2015uus}), yielding
\begin{align} \label{eq:soft}
  S_{ijk}^\one(b_x, \eta_J, \mu, \nu) &= - \sum_{i<j} {\bf T}_i \sdt {\bf T}_j S^\one\Bigl(b_x,\mu, \nu \sqrt{n_i \sdt n_j/2}\Bigr)
  \,,\nn \\
  S_{ijk}^\two(b_x, \eta_J, \mu, \nu) &= - \sum_{i<j} {\bf T}_i \sdt {\bf T}_j S^\two\Bigl(b_x,\mu, \nu \sqrt{n_i \sdt n_j/2}\Bigr)
  \nn \\ & \quad
  +\frac12\Bigl[S_{ijk}^\one(b_x,\eta_J,\mu, \nu)\Bigr]^2
\,,\end{align}
where $S = 1 + \alpha_s/(4\pi) S^{(1)} + \dots$ with
\begin{align}
  S^{(1)}(\vec b_\perp,\mu, \nu)
  = - 2 L_b^2 + 8 L_b \ln \frac{\mu}{\nu} - \frac{\pi^2}{3}
\,,\end{align}
and $L_b = \ln(\vec b_\perp^{\,2}\mu^2 e^{2\ga_E}/4)$.
The color factors are ${\bf T}_q \sdt {\bf T}_{\bar q} = \tfrac16$ and ${\bf T}_q \sdt {\bf T}_g = {\bf T}_{\bar q} \sdt {\bf T}_g = -\tfrac32$, and $n_a \sdt n_b = 2$, $n_{a,b} \sdt n_J = 1 \mp \tanh \eta_J$.
 The contribution involving exchanges between three Wilson lines vanishes due to color conservation~\cite{Becher:2012za}. 

The beam functions describe the transverse momentum of the colliding hard parton with respect to the beam axis due to collinear initial-state radiation. They have a perturbative matching onto PDFs, and the matching coefficients are
known at two-loop~\cite{Catani:2011kr,Catani:2012qa,Gehrmann:2012ze,Gehrmann:2014yya,Luebbert:2016itl,Echevarria:2016scs,Luo:2019hmp,Luo:2019bmw} and three-loop order~\cite{Behring:2019quf,Luo:2019szz,Ebert:2020yqt}. In \eq{fact}, we only probe the $x$-component of the transverse momentum. As azimuthal symmetry is broken, we get a  linearly-polarized contribution, which encodes the effect of a spin-superposition of the gluon extracted from the proton. This linear polarization was not taken into account in previous studies of the azimuthal angular decorrelation.

The TMD jet function describes the offset of the WTA axis with respect to the jet momentum (for the standard jet axis there is no transverse momentum dependence).
We recalculated the TMD jet functions~\cite{Gutierrez-Reyes:2018qez,Gutierrez-Reyes:2019vbx} using the $\eta$-regulator, taking $\de \phi \ll R$, which removes all dependence on the jet radius. In this limit the momentum of the initial parton is contained in the jet, which simplifies its expression. Writing $\cJ_i = 1 + \al_s/(4\pi) \cJ_i^\one + \dots$, 
\begin{align}
    \cJ_q^{(1)}(\vec b_\perp,\mu,\nu) & = C_F \Bigl[ L_b \Bigl( 3 + 4\ln\frac{\nu}{\omega} \Bigr) + 7 -\frac{2\pi^2}{3} - 6\ln 2 \Bigr]\,, \nn \\
    \cJ_{g}^{(1)}(\vec b_\perp,\mu,\nu) & = C_A \Bigl[ L_b \Bigl( \frac{11}{3} \!+\! 4\ln\frac{\nu}{\omega} \Bigr) \!+\! \frac{131}{18}  \!-\! \frac{2\pi^2}{3} \!-\! \frac{22}{3}\ln 2 \Bigr]
    \nn \\ & \quad
    + T_F n_f \Bigl[ -\frac{4}{3}L_b - \frac{17}{9} + \frac{8}{3}\ln 2\Bigr]\,,
\end{align}
where $\w = 2 p_T^J \cosh \eta_J$. Here $b_\perp$ is transverse to the jet axis, and in \eq{fact} we take it also perpendicular to the beams with $|\vec b_\perp| = |b_x|$. \\

{\it Linearly-polarized gluon jet function. --}
The linearly-polarized jet function describes the effect of a spin-superposition of the gluon initiating the jet, and is defined as
\begin{align} \label{eq:J_linear}
&\cJ_{g}^L(\vec b_\perp, \mu, \nu) 
\\ & \quad
= \biggl[\frac{1}{d-3} \Bigl(\frac{g_\perp^{\mu\nu}}{d-2} + \frac{b_\perp^\mu b_\perp^\nu}{\vec b_\perp^{\,2}}\Bigr)\biggr] \frac{2(2\pi)^{d-1} \w}{N_c^2-1}
\nn \\ & \qquad \times
\langle 0| \delta(\w - \bar n \sdt {\mathcal P}) \delta^{d-2}({\mathcal P}_\perp) {\mathcal B}_{n\perp\mu}^{a}(0)  e^{\img \vec b_\perp \cdot {\hat {\vec k}}_\perp}  {\mathcal B}_{n\perp\nu}^{a}(0) |0\rangle 
\,.\nn \end{align}
It differs from the standard jet function by the factor in square brackets (which is otherwise $-g_\perp^{\mu\nu}/(d-2)$).
Taking $\vec n_J$ to indicate the jet direction, we introduce light-cone vectors $n^\mu = (1, \vec n_J)$ and $\bn^\mu = (1, -\vec n_J)$, with $\perp$ denoting components transverse to both. ${\mathcal B}_{n\perp\mu}^{a}$ is the collinear gluon field, which includes a collinear Wilson line to ensure gauge invariance. The initial momentum of the field is fixed by the delta functions. The transverse momentum corresponding to the displacement of the WTA axis with respect to this initial momentum is picked out by ${\hat {\vec k}}_\perp$. The first non-vanishing order of \eq{J_linear} is one loop, for which we obtain 
\begin{align}
  \cJ_{g}^{L\one}(\vec b_\perp, \mu, \nu) =  - \frac{1}{3} C_A  + \frac{2}{3} T_F n_f
\,.\end{align}
Since this is the first nontrivial order, it yields the same result for other recoil-insensitive axes.

We provide evidence for contributions from linearly-polarized gluon beam and jet functions in Fig.~\ref{fig:mcfm}, by showing the difference between the cross section obtained using our factorization in \eq{fact} and \MCFM at NLO \cite{Campbell:2002tg,Campbell:2003hd}, with a cut $\de \phi < \de \phi^{\rm cut}$.  
This difference should vanish in the limit $\de \phi^{\rm cut} \to 0$, but only does so when the linearly-polarized gluon beam and jet functions are included. 
(Note that the linearly-polarized contributions are not visible in the cross section differential in $\de \phi$.)
The left panel shows the contribution involving $qq$ PDFs, which only involves linearly-polarized beam functions, and in the right panel we focus on the $n_f$ dependent contribution from $q\bar q$ PDFs, to provide evidence for 
a nonzero contribution from linearly-polarized jet functions.

{\it Track-based measurement. --}
\begin{figure}
\includegraphics[width=0.45\textwidth]{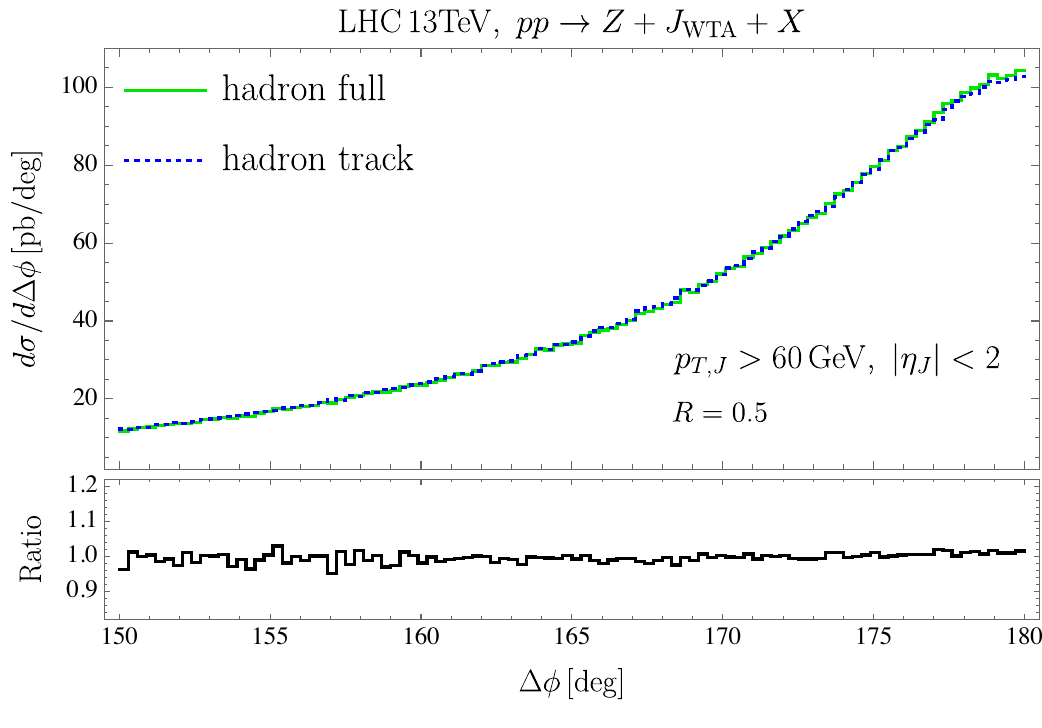}
\caption{Predictions from \Pythia for the azimuthal angle between the vector boson and jet, using all particles (green) or only charged particles (blue dotted).
\label{fig:pythia}}
\end{figure} 
The angular resolution of jet measurements is about 0.1 radians, due the size of
the calorimeter cells, limiting access to the resummation region. This can be overcome by measuring the jet using only charged particles, exploiting the superior angular resolution of the tracking systems at the LHC. Here we identify another advantage of the WTA axis: since the effect of soft radiation on the jet algorithm is power suppressed, switching to a track-based measurement only modifies the jet function. (Note that $p_{T,J}$ and $\eta_J$ do not require a fine angular resolution and are therefore measured on the full jet.) Consistency of the factorization formula in \eq{fact} then implies that this track-based jet function $\bar \cJ$ has the same anomalous dimension as $\cJ$. We reach the same conclusion by a direct calculation using track functions~\cite{Chang:2013rca,Chang:2013iba}. Explicitly, the difference in the one-loop constant for the quark jet function is
\begin{align}
   \bar \cJ_q^\one &= \cJ_q^\one  + 
   4C_F \int_0^1\! \df x\, \frac{1+x^2}{1-x} \ln \frac{x}{1-x} \int_0^1\! \df z_1\, T_q(z_1,\mu)
   \nn \\ & \quad \times
   \int_0^1\! \df z_2\, T_g(z_2,\mu) [\theta(z_1 x - z_2 (1-x)) - \theta(x-\tfrac12)]
\,,\end{align}
in terms of the track functions $T_i(z,\mu)$. The change reflects the
possibility of a hadronization mismatch in the WTA recombination: A losing
(in WTA sense) parton may hadronize into the winning track. The expression for
the gluon jet function involves the appropriate replacement of the splitting functions, and there is no modification to the linearly-polarized gluon jet function at order $\al_s$. We have verified using \Pythia 8.2~\cite{Sjostrand:2014zea} that using tracks only has a minimal effect on this measurement, see \fig{pythia}. For the standard jet axis, this difference is larger~\cite{longpaper}. The conclusions reached here also apply to other angular measurements, such as in~\cite{Gutierrez-Reyes:2018qez,Gutierrez-Reyes:2019vbx,Gao:2019ojf}. Recently, the ease of including track functions for purely collinear measurements was demonstrated~\cite{Chen:2020vvp}.

\begin{figure}
\includegraphics[width=0.45\textwidth]{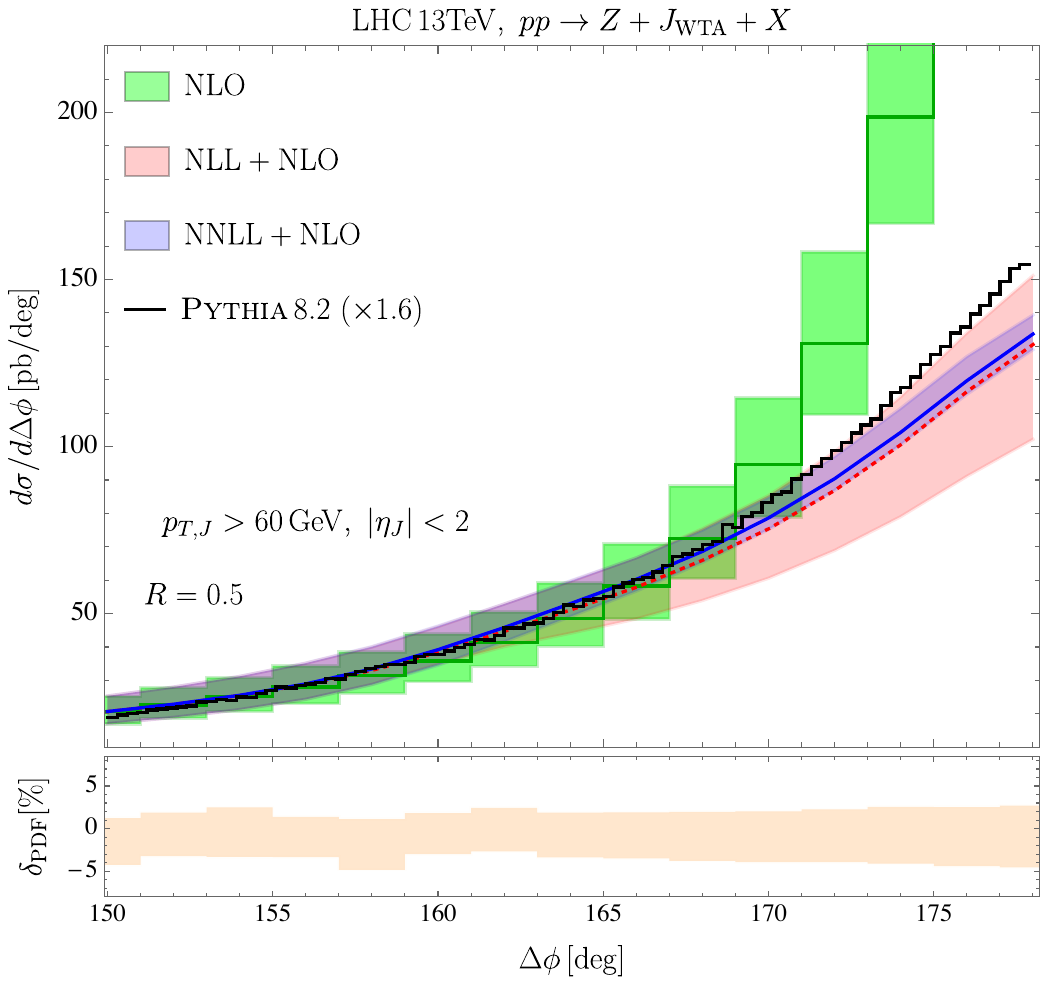}
\caption{Our resummed prediction for the cross section differential in the azimuthal angle at NLL+NLO (red) and NNLL+NLO (blue), compared to the NLO cross section from \MCFM (green) and \Pythia at hadron level (black). The lower panel shows the PDF uncertainty.
\label{fig:resummation}}
\end{figure}

{\it Resummed predictions. --}
We obtain predictions in \fig{resummation} for the LHC with $\sqrt{s}=13$~TeV, using the factorization formula in \eq{fact}. Jets are identified by the anti-$k_T$ clustering algorithm with $R=0.5$ and the WTA recombination scheme. We use the CT14nlo parton distribution functions~\cite{Dulat:2015mca}, and show the PDF uncertainty.

We show our resummed predictions in \fig{resummation} at NLL+NLO and NNLL+NLO
order, and compare to the NLO cross section obtained from \MCFM. For our central
curve we take $\mu_H = \sqrt{p_{T,V}^2+m_V^2}$, $\nu_S = \mu_B= 2
e^{-\gamma_E}/|b_x|$, $\nu_{B_{a,b}}=x_{a,b}\sqrt{s}$ and $\nu_J = \omega$. We
estimate the perturbative uncertainty by varying $\mu_B$ and $\mu_H$ by a factor
two around their central values, taking the envelope of the scale variations. The uncertainty bands of the NLL and NNLL predictions overlap, and are substantially reduced for the NNLL result in the resummation region $\Delta \phi \gtrsim 170^{\circ}$. While the resummed predictions approaches a constant in the back-to-back limit, the NLO prediction becomes unreliable due to unresummed logarithms. At very low values of $p_{x,V}$, the scale $\mu_B$ hits the Landau pole. To avoid this unphysical behaviour, we apply the $b^*$-prescription $|b_x|\to b^*=|b_x|/\sqrt{1+b_x^2/b_{\rm max}^2}$ \cite{Collins:1984kg}. On the other hand, for $\Delta \phi \lesssim 160^{\circ}$ the fixed-order corrections are important. These are included by matching to the NLO using a transition function, as in e.g.~\cite{Becher:2019bnm}. 

We also compare to \Pythia, including the NLO $K$-factor of 1.6. The difference in shape for $\Delta \phi\gtrsim 170^{\circ}$ is not significant, given the size of the NLL uncertainty band (a reasonable proxy for the \Pythia uncertainty). We have verified that this is not due to multiparton interactions or hadronization effects, which have a minimal effect on this observable.

{\it Conclusions. --}
In this letter we establish a promising avenue for precision studies of transverse momentum distributions in initial and final states of high energy collisions. We present the first prediction of the azimuthal angular distribution in boson-jet production at NNLL accuracy. Such high theoretical precision is achieved by the use of a recoil-free jet axis in the azimuthal angle definition, ensuring that non-global logarithms are absent. Using simulations at truth particle level, we demonstrate that measuring this angle using charged tracks accurately reproduces the distribution determined when using all the jet particles. This allows us to exploit the finest angular resolution for precise experimental measurements.

Our theoretical predictions are based on a factorized expression in SCET involving TMD beam and jet functions. This factorization is confirmed (beyond consistency of anomalous dimensions) at NLO by comparing to \MCFM, verifying the necessity of including linearly-polarized TMD distributions. These functions are intimately related to TMD PDFs and fragmentation functions in the literature, describing the nonperturbative regime $| b_x| \sim 1/\Lambda_{\rm QCD}$, where the perturbative matching onto collinear PDFs fails. However, the soft function is different compared to e.g.~Drell-Yan, and so one cannot simply absorb it into the TMD parton distribution, as is customary~\cite{Collins:2011zzd,GarciaEchevarria:2011rb} (this is e.g.~precluded because our soft function depends on the jet rapidity).

Our work serves as a baseline for pinning down the inner workings of the QCD medium produced in heavy-ion collisions \cite{Chen:2018fqu}, where the use of a recoil-free axis will be even more important to suppress effects from the huge underlying event background. This study also presents an excellent opportunity to shed light on the three-dimensional picture of gluon dynamics inside the proton. Furthermore, polarization effects from initial and final states can also be included in our framework. We are confident that these considerations offer ample opportunities for future research. 

{\it Acknowledgements --}
We thank Iain Moult, Frank Tackmann and Feng Yuan for discussions and comments on the manuscript.  This work is supported by the ERC grant ERC-STG-2015-677323, the NWO projectruimte 680-91-122, the National Science Foundation under Grant No.~PHY-1720486 and PHY-1915093, Center for Frontiers in Nuclear Science of Stony Brook University and Brookhaven National Laboratory, and the D-ITP consortium, a program of NWO that is funded by the Dutch Ministry of Education, Culture and Science (OCW). 

%

\end{document}